\documentclass[runningheads]{llncs}
\usepackage[T1]{fontenc}
\usepackage{graphicx}
\usepackage{todonotes}
\usepackage{comment}
\usepackage[table]{xcolor}
\usepackage{longtable}
\usepackage{array}
\usepackage[margin=1in]{geometry}
\usepackage{pdfpages}
\usepackage{hyperref}
\usepackage{enumitem}
\usepackage{breakcites}

\definecolor{headerbg}{RGB}{220,220,220} 
\definecolor{DCcolor}{RGB}{255,235,235}    
\definecolor{PCcolor}{RGB}{235,235,255}    
\definecolor{groupbg}{RGB}{255,255,204}    

\begin{document}
\title{Data-Related Challenges and Requirements for Event Log Generation in Process Mining: A Systematic Literature Review}
\titlerunning{Data-related Challenges and Requirements for Event Log Generation}
%

\author{Ghita El Alaoui Talibi \inst{1}\thanks{This work was completed while the author was affiliated with the institution.} \and
Oleksandr Kosenkov\inst{2,3}\orcidID{0000-0002-9971-1130} \and Anastasija Nikiforova
\inst{4}\orcidID{0000-0002-0532-3488}}
\authorrunning{G. Talibi et al.}

\institute{Technical University of Munich, Boltzmannstraße 3, 85748 Garching bei München
\and
fortiss GmbH, Guerickestraße 25, 80805 München\\
\and
Blekinge Institute of Technology, Valhallavägen 10, 371 79 Karlskrona, Sweden\\
\email{\{firstname.lastname\}@bth.se}
\and
University of Tartu, Ülikooli tn 18, 50090 Tartu, Estonia\\
\email{\{firstname.lastname\}@ut.ee}\\
}

\maketitle              
\begin{abstract}
Process mining is becoming an essential technology that not only supports the improvement of business processes but also enables the software technologies underlying the future digital economy. Among the multiple challenges discussed by the process mining community, data quality is emerging as the most pressing one. However, insufficient attention has been paid to event log generation as the very first phase of process mining and to its relationship with data engineering and requirements engineering. To address this gap, we conducted a systematic literature review of data-related challenges, requirements, and solutions in event log generation. Our results show that a wide range of 29 challenges can be addressed with 5 basic categories of requirements. Existing solutions already embed some form of data engineering and domain knowledge engineering techniques for event log generation. This review can support researchers and practitioners in understanding current trends in event log generation for process mining and in leveraging data and requirements engineering techniques to improve log generation and, ultimately, process mining outcomes.

\end{abstract}

\section{Introduction}
Process mining is increasingly recognized as a key enabler of value creation in digitally-enabled businesses ~\cite{badakhshan2022creating} and as a supporting technology for emerging technological developments such as robotic process automation~\cite{el2023robotic}, internet of things~\cite{singh2022using}, digital twins~\cite{grobis2025classification}, and industry 4.0~\cite{osman2019industry}. In practice, process mining is applied for purposes such as compliance assessment, audit support, performance analysis and business intelligence.~\cite{vom2021five}. Logging plays a critical role in software system development and operations ~\cite{bogatinovski2022qulog}, providing the primary data source for understanding and improving software-related processes. In this context, process mining offers a promising means to further contribute to software engineering practice by enabling data-driven analysis of software processes. The wide range of domains in which process mining has been applied highlights its importance for the development of emerging digital technologies and the digital economy overall. However, despite its potential, several challenges and the relative novelty of process mining research and practice continue to limit its widespread adoption. In particular, data quality plays a fundamental role in the quality of process mining resulting insights, and data-related challenges remain among the most critical to address \cite{ter2023process}. Moreover, the increasing adoption of Artificial Intelligence (AI) expands data processing opportunities but simultaneously makes data quality issues in process mining even more critical.

According to the process mining manifesto, one of the core guiding principles is that event data should be treated as \textit{first-class citizens}, i.e., as playing a fundamental role in process mining and therefore requiring corresponding attention. The manifesto further identifies the discovery, integration, and cleaning of event data as the most important challenges in process mining, effectively pointing out that they remain largely unresolved~\cite{van2012process}. Despite the recognized importance of data, data-related challenges remain unresolved ~\cite{ter2023process}. In recent years, several process mining studies have proposed solutions to address data quality issues in event log generation~\cite{montali_bermuda_2023,andrews_quality-informed_2020}. However, there is no overview of the state of research that covers both data-related challenges and the specific requirements needed to address them. 

This lack of a systematic mapping between data-related challenges and the requirements for addressing them limits both researchers and practitioners in identifying critical research gaps, prioritizing improvement efforts, and designing effective solutions for event log generation.

To address these issues, this study conducts a systematic literature review to understand persistent data-related challenges, the requirements for addressing them, and corresponding solutions. Our study contributes to the existing state of research in three key ways. 
\begin{enumerate}[nosep]
    \item Systematic Synthesis. As the body of process mining research is growing, it is essential to synthesize existing results systematically. We provide a systematic overview that should help both researchers and practitioners navigate the most important topics and easily identify challenges in event log generation requiring close attention or receiving insufficient attention, requirements that need to be implemented first of all to address particular groups of challenges.
    \item Focus on Event Log Generation. While much of current research focuses on cleaning and refining event logs after they are generated ("post-hoc"), we argue that a better approach is to address data quality challenges as early as possible, during the event log generation process itself. However, this proactive shift is constrained by a persistent divide in research: logging (data generation) and log analysis (process mining) are typically studied in isolation, preventing a holistic understanding of their intricate interdependencies~\cite{gu2022logging}. To enable high-quality, process-aware logging from the start, it is essential to bridge this gap by focusing on the intersection of log generation and log analysis, ensuring that logging systems are designed with the requirements of process mining in mind. We hence focus on data-related challenges and requirements in event log generation.
    \item Interdisciplinary Approach (Data and Requirements Engineering). Process mining is a relatively recent, however, comprehensive discipline, reusing the methods and approaches of other disciplines such as data mining, without often specifically accounting for such intersections. We look into the intersection of process mining with data engineering and requirements engineering to facilitate more effective synergies of process mining with these two disciplines. Data engineering research and practice also deal with the quality of the data and the corresponding methods to address data-related challenges. The potential role of requirements engineering is twofold. It can help to establish the data-related requirements for process mining purposes. Also, requirements engineering can further contribute by formulating logging requirements for software engineering to ensure that logs generated by software systems are suitable for process mining. Taken together, these disciplines may contribute to addressing the data-related challenges and requirements with their existing methods to facilitate better results in process mining.
\end{enumerate}

From a practical perspective, we seek to support the development of log generation methods and software systems that would facilitate the application of process mining and its capabilities through synthesizing an overview of corresponding challenges, requirements, solutions, and describing the potential contribution of data and requirements engineering.
To achieve the set goals, we conduct a systematic literature review to understand persistent data-related challenges, requirements for addressing challenges, and corresponding solutions. We also identify the existing gaps in process mining research and discuss how data engineering and requirements engineering techniques contribute in event log generation.

The rest of the paper is structured as follows. In the following section~\ref{sec:relatedWork}, we provide an overview of related work. In the Background section~\ref{sec:background}, we introduce the basic concepts used in this paper. We describe the methodology we have followed to execute our study in section~\ref{sec:methodology}. In section~\ref{sec:results}, we present our results of the literature review focusing on data-related challenges, requirements to address these challenges, and existing solutions in corresponding subsections. We discuss our results in section~\ref{sec:discussion}. Our approach to address threats to validity is described in section~\ref{sec:threatsValidity}, and we conclude our study with section~\ref{sec:conclusion}. The open data set for this study is available in the following \href{https://zenodo.org/records/18518836}{Zenodo repository (DOI: 10.5281/zenodo.18518836)}.

\section{Related work}\label{sec:relatedWork}
The recent study~\cite{ter2023process} identified process-data quality as a key frontier in process mining and examined several fundamental data-quality issues, including how to determine whether the right data are being collected and how to prevent process-data quality problems. While this study highlights the importance of data quality, it does not provide a systematic mapping between specific data-related challenges and requirements, leaving practitioners without actionable guidance for implementing quality-focused logging systems.

\cite{diba_extraction_2020} provided an overview of techniques on extraction, correlation, and abstraction of event data. Despite being a valuable source of information this study did not followed a systematic methodology to select the studies for an overview. Also, this study mainly focused on available solution techniques rather than the underlying challenges these methods aim to address.

\cite{kampik_event_2022} conducted a survey to identify the main conceptual and technical challenges in event log generation in practice. However, the study explores only a subset of challenges and does not provide a systematic categorization.

In 2015 \cite{van2015extracting} provided 12 guidelines for logging. However, these were not derived empirically, were not connected to specific challenges, and now are at least partially outdated in light of recent process mining advancements.

\cite{fahland_event_2020} reported challenges and corresponding solutions for event log generation in healthcare system. The study categorized the challenges according to the steps of log generation, however, have not suggested any higher-level synthesis or cross-domain perspective.

We suggest that in the future more systematic overview of challenges and requirements to address them would benefit future empirical studies like \cite{fahland_event_2020} by providing them a baseline for more systematic reporting.

Overall, to the best of our knowledge, no existing study provides a systematic review that explicitly focuses on data-related challenges and requirements in event log generation and, importantly, establishes a structured mapping between them. While prior work discusses individual challenges and, in some cases, corresponding requirements, these aspects are typically addressed in isolation, making it difficult to understand their interrelationships. Moreover, many studies exhibit inconsistencies in how challenges and requirements are presented—for example, by outlining high-level challenges while introducing more specific, yet disconnected, requirements, or by focusing on only one of these aspects without explicitly linking it to the other. Also, connection of process mining to data engineering and requirements engineering stays understudied.
Overall, these studies provide insights into methods, guidelines or domain-specific challenges, but they do not provide a systematic overview of data-related challenges and requirements. Without such an overview, research efforts remain fragmented, with isolated challenges and solutions repeatedly studied without a shared frame of reference. At the same time, practitioners lack guidance for prioritizing logging and data quality improvements that most strongly impact process mining outcomes, limiting the reliability and practical adoption of process mining. The lack of an integrated perspective limits the opportunities to effectively leverage established practices from data engineering and requirements engineering, reinforcing reliance on costly post-hoc event log refinement rather than supporting proactive, quality-focused logging at the point of data generation.

\section{Background}\label{sec:background}
Logging in software and information systems can be conducted for multiple purposes such as tracing the correctness of behavior of the system and particular system events~\cite{gu2022logging}, debugging, failure handling, system analysis, system recovery~\cite{rong2017systematic}. However, logging also can be an important source of information about business process related events executed by the system. This way, it is important to distinguish between system event logs which are focused on recording the system behavior and \textit{business event logs} which are aimed at recording the information relevant for the execution of business processes. Usually (business) event logs are not readily available from the system and they need to be generated with the application of different techniques. Raw event data generated by software systems and system event logs often lack the structure and context needed for direct process mining analysis. It is transformed into \textit{business event logs}, which are structured representations of process executions containing sufficient information about business process execution. In this study, we use the term \textit{event log} as a synonym of business event log, as system event logs are out of scope of our study. Business event logs are an important input for \textit{process mining} which is the application of data science and mining techniques for the discovery, analysis, monitoring, improvement of business processes. We define \textit{event log generation} as processing of raw event data and system logs for obtaining business event logs applicable for process mining purposes.

\textit{Data engineering} is a software engineering practice that involves the end-to-end management of data throughout its lifecycle, encompassing data ingestion, storage, integration, modeling, preprocessing, processing, and querying, to enable data to be prepared and used for analysis.~\cite{romero_dataengineering_2020}

\textit{Requirements engineering} is a branch of software engineering focusing on the systematic elicitation, modeling, and analysis of stakeholders’ needs and constraints, as well as relevant domain information, in order to define and achieve the goals of a software system.~\cite{nuseibeh_requirementseng}

In the context of this study, \textit{data-related challenges} are challenges related to processing data, which emerge in the process of event log generation for process mining.
And \textit{data-related requirements} are requirements that need to be implemented in the process of event log generation for process mining.

\section{Methodology}\label{sec:methodology}
To execute this study we have conducted a systematic literature review  following the guidelines by Kitchenham~\cite{Kitchenham2007guidelines}. We have also followed empirical standards for software engineering research~\cite{ralph2020empirical}, while designing and executing the study to assure the quality of the research, and address the potential threats to validity. Our study can be characterized as meta-synthesis~\cite{ralph2022paving} aiming at (1) providing a systematic overview of data-related challenges and requirements in event log generation for process mining, and (2) identifying the relevance of data engineering and requirements engineering techniques to address the challenges and requirements. In order to achieve the two aforementioned goals we have formulated the following three questions:
\begin{enumerate}[nosep]
    \item[RQ1:] What are the data-related challenges in event log generation for process mining, as reported in the literature?
    \item[RQ2:] What are the data-related requirements for event log generation for process mining, as identified in the literature?
    \item[RQ3:] How are existing data engineering and requirements engineering techniques applied to address the identified data-related challenges?
\end{enumerate}

Next, we report each step of the systematic literature review.

We limit our search to the last 5 years to cover the most recent advancements in process mining. Although process mining research dates back to the early 2000s, widespread commercial adoption ~\cite{appsruntheworld2025process}  and significant market growth of commercial process mining solutions emerged primarily in the late 2010s (2018-2019) ~\cite{venturebeat2021why}, making recent studies more representative of current practices and challenges. 

\subsection{Search strategy}
Taking into account that we have not identified any comprehensive existing studies related to the goals of our study, we have selected a search-based strategy for the identification of primary studies. We have selected Scopus and IEEE Xplore databases for this. We selected Scopus due to its comprehensive coverage of primary studies. Additionally, we included IEEE Xplore, as recent reports indicate that its search results have the least overlap with Scopus compared to the ACM Digital Library and other databases~\cite{valente2022analysis}.

In order to develop the research query, we have used an initial set of key phrases centered around the studied phenomena, such as “event log generation”, “event log extraction” and associated synonyms (e.g., "event", "logging"). During the trial searches, we have also applied different filters and searched across full texts as well as titles, abstracts and keywords. Some of the tried queries returned too few studies, while others were too broad and yielded many irrelevant studies.

We validated the effectiveness of the search query against key benchmark studies we found essential for our research, namely~\cite{andrews_quality-informed_2020,kampik_event_2022,de_weerdt_foundations_2022,fahland_event_2020,diba_extraction_2020}.

For some of the tried search queries we have also conducted a trial backward snowballing to identify additional relevant studies. During this process, the inclusion of broader terms such as "event data", "extraction", and "logging" was recognized as important. The query was iteratively updated based on the screening and snowballing and validation against key papers. Eventually, we defined the following search query:

\textit{(“event log” OR “event data”) AND (“generation” OR “extraction” OR “logging”) AND “process mining”}

This query was defined for its ability to capture a comprehensive range of studies related to event log generation, to include synonyms and variations in terminology to ensure no relevant studies were missed, and to focus specifically on process mining, excluding irrelevant studies from unrelated domains. It is aligned with the broad definition of event log generation and research goals. We also executed a trial backward snowballing using the final search query, this however, has not yielded a significant number of new papers. Due to this the decision was made not to conduct snowballing. The final selected query included all of the benchmark studies.

\subsection{Study selection and screening}
The following criteria were applied during the screening and selection process, initially based on titles and abstracts, and subsequently on full-text review.
\subsubsection{Inclusion Criteria}
\begin{itemize}[nosep]
    \item \textbf{IC01} Study is in English.
    \item \textbf{IC02} Study discusses event data logging systems, practices, or standards relevant to process mining.
    \item \textbf{IC03} Study addresses data quality challenges during logging, extraction, or transformation of event logging data.
    \item \textbf{IC04} Study focuses on event log generation, including raw data capture, preparation, or structuring.
    \item \textbf{IC05} Study provides insights into data requirements, methodologies,  tools, or data engineering techniques related to event log generation.
\end{itemize}

\subsubsection{Exclusion Criteria}
\begin{itemize}[nosep]
    \item \textbf{EC01} Study is not published in a peer-reviewed scientific journal, conference, or workshop. This includes technical reports, guidelines, project deliverables, theses, extended abstracts, and books.
    \item \textbf{EC02} Study focuses on event log refinement or repair.
    \item \textbf{EC03} Study is about model-based event log generation or model extraction from event logs.
    \item \textbf {EC04} Study on synthetic log generation or unrelated logging techniques.
\end{itemize}

\subsubsection{Search and Screening Process}
Next, we provide a short overview of the main steps of the systematic literature review with the number of primary studies processed at each step.
\begin{itemize}[nosep]
    \item The initial search retrieved 561 studies: 376 from Scopus and 185 from IEEE Xplore. The search was conducted without applying any publication time filter.
    \item During duplicate removal, 82 studies were excluded, leaving 479 unique studies.
    \item After screening the remaining studies, all papers published before 2019 were excluded (169 studies), resulting in 310 primary studies.
    \item Following title and abstract screening, 191 studies were excluded, leaving 119 papers for full-text review.
    \item The full texts of these 119 papers were reviewed, and 59 papers were ultimately selected for data extraction.
\end{itemize}

After identifying the final set of articles from the study selection process, a systematic data extraction protocol was established and implemented to capture the key elements necessary to address our research questions. This process was designed to be transparent and replicable and was based on an Excel-based data extraction form with clearly defined categories. Each selected paper was read in full, and relevant details were recorded manually.

To systematically document the findings, we used an Excel spreadsheet as our central repository, designing a template with the following predefined columns linked to the following research questions (RQ):

 \begin{itemize}[nosep]
        \item \textbf{Metadata/ Bibliographic Details:} Title, Authors and Year to maintain accurate records and document each study for traceability and for later cross-referencing.
        \item \textbf{Challenges (RQ1)}: Common issues, recurring problems and challenges for high-quality event log generation. This category directly addresses RQ1 by identifying the data-related challenges in event log generation.
        \item \textbf{Requirements (RQ2)}: Data-related requirements for effective high-quality event log generation. It could be related to data structure and format requirements, data quality considerations, data assumptions or constraints, or even data-related tasks as they are defining what makes the data "ready" for use. This category addresses RQ2 by detailing the data-related requirements identified in the literature.
        \item \textbf{Solutions (RQ3)}: This category addresses RQ3 by identifying the proposed methods, tools, or frameworks to overcome the identified challenges and meet the requirements for event log generation.
    \end{itemize}

Among the primary studies we have identified some studies that reviewed multiple previous studies (e.g., \cite{diba_extraction_2020}). Although these studies did not follow a systematic review methodology, they discussed relevant techniques and were therefore included in our analysis. We also performed an ad hoc quality check by screening for publication types typically associated with lower methodological rigor, such as research previews, doctoral symposium papers, short papers, technical reports, white papers, and other non–peer-reviewed publications. However, no such studies were identified in the final set. The study selection and data extraction sheets used in this study are publicly available on \href{https://zenodo.org/records/18518836}{Zenodo (DOI: 10.5281/zenodo.18518836)}.

\subsubsection{Data analysis and synthesis}
For data analysis, an additional column was added for the coding of each category, namely challenges (RQ1), requirements (RQ2), solutions (RQ3). Initially, our extraction form included a separate Data Quality category (RQ1, RQ2, RQ3) intended to capture any explicit metrics, dimensions, or frameworks used to ensure data quality. However, upon reviewing the studies, we found that only a few papers provided distinct details on data quality. As a result, these sparse mentions were integrated into the Requirements category, which offered a more comprehensive context for understanding the conditions and constraints necessary for effective event log generation.

In the first cycle of analysis, open coding was employed to create an extensive list of descriptive labels. In the first pass of labeling, the intention was to remain as faithful as possible to the language of the studies. Once all extracted data were given an initial label, a second pass was conducted to harmonize the terminology and refine any inconsistencies. In this second iteration, similar or overlapping codes were merged into unified categories. Following this two-round labeling, for each category (challenges, requirements, solutions) a separate reference mapping table was maintained in spreadsheets to document which papers were tackling each label. Throughout these iterations, reference IDs of the papers from which each label originated were maintained in a separate sheet, enabling traceability. For our mapping of challenges to requirements (Tab.~\ref{tab:table}), we relied on the results of our challenge analysis and established the mapping based on the authors’ interpretation of the relationships, rather than solely on explicit links reported in the primary studies. This interpretative approach was necessary due to the lack of a consistent and explicit discussion of the interconnections between challenges and requirements across the reviewed literature. In many primary studies, challenges are presented primarily as general motivation and are not directly linked to the corresponding requirements, or they are discussed at different levels of abstraction. To answer RQ3 we identified the solutions are data engineering or requirements engineering related.

\section{Results}\label{sec:results}
\subsection{Data-related challenges in event log generation for process mining (RQ1)}
The analysis reveals a wide range of 29 data-related challenges in event log generation for process mining. The most frequently cited challenges include data quality issues (39\%), data integration (27\%), and non-process centric data (24\%). Next, we provide an overview and short description of the eight challenges mentioned in ten or more primary studies for an overview of all 29 types of challenges see Table~\ref{tab:table}.

\textbf{Ch1: Data quality} issues were mentioned in 23 primary studies (39\%). This encompasses a range of issues, including missing data (12 references, 20\%)(e.g., ~\cite{de_weerdt_foundations_2022,fahland_event_2020,munoz-gama_rethinking_2022,umer_data_2022,prostean_event_2020,van_der_aalst_practitioners_2022,sim_automatic_2022,butt_behavioral_2023,montali_bermuda_2023,geeganage_text2el_2022,andrews_quality-informed_2020,desjardins_enabling_2024}, imprecise data (6 references, 10\%)\cite{de_weerdt_foundations_2022,geeganage_text2el_2022,van_der_aalst_practitioners_2022,augusto_process_2022,olveczky_object-centric_2019,andrews_leveraging_2019}, inconsistent data (9 references, 15\%)\cite{fahland_event_2020,munoz-gama_rethinking_2022,umer_data_2022,desjardins_enabling_2024,van_der_aalst_federated_2021,van_der_aalst_practitioners_2022,de_smedt_llms_2024,montali_bermuda_2023,geeganage_text2el_2022}, incorrect data (5 references, 8\%)\cite{de_weerdt_foundations_2022,umer_data_2022,montali_bermuda_2023,geeganage_text2el_2022,andrews_quality-informed_2020} and irrelevant data (3 references, 5\%)\cite{de_weerdt_foundations_2022,umer_data_2022,fahland_event_2020}. The literature emphasizes that data quality issues can significantly impact the results of process mining analyses.

\textbf{Ch2: Data integration} from multiple sources into a coherent event log was mentioned as a challenge in 16 primary studies (27\%) \cite{kampik_event_2022,diba_extraction_2020,de_weerdt_foundations_2022,fahland_event_2020,schuh_data_2020,umer_data_2022,prostean_event_2020,seiger_towards_2020,desjardins_enabling_2024,van_der_aalst_practitioners_2022,munoz-gama_event_2022,milde_enabling_2023,van_der_aalst_federated_2021,montali_event_2023,martinez_lagunas_analysis_2024,tariq_time_2022}. The literature notes that data required for event logs often resides in various systems, and integrating this data requires significant effort, particularly when dealing with non-process-centric data.

\textbf{Ch3: Non-process centric data} emerged as the challenge mentioned in 14 primary studies (24\%) \cite{diba_extraction_2020,de_weerdt_foundations_2022,marrella_towards_2022,umer_data_2022,alzhrani_process-aware_2024,prostean_event_2020,seiger_towards_2020,benvenuti_interactive_2022,desjardins_enabling_2024,martinez_lagunas_analysis_2024,breitmayer_towards_2022,morichetta_event_2025,di_francescomarino_extracting_2019}. This indicates that data not inherently designed for process mining poses significant difficulties. For example, data stored in relational databases or enterprise systems often lacks a process-oriented structure, making it challenging to extract meaningful event logs.

\textbf{Ch4: Heterogeneity} in data sources and structures was mentioned in 13 primary studies (22\%) \cite{kampik_event_2022,diba_extraction_2020,marrella_towards_2022,bekeneva_approach_2020,umer_data_2022,hernandez_semi-automated_2022,benvenuti_interactive_2022,just_collaborative_2023,de_smedt_llms_2024,gonzalezlopez_connecting_2019,moctar_mbaba_blockchain_2022,milde_enabling_2023,valencia-parra_enabling_2019}. This underscores the difficulty in integrating and interpreting data from diverse sources. The literature highlights that event logs often need to be constructed from data scattered across multiple systems, each with its own format and structure.

\textbf{Ch5: Case notion} was the most cited challenge, with 13 references (22\%)\cite{kampik_event_2022,de_weerdt_foundations_2022,fahland_event_2020,umer_data_2022,indulska_extracting_2023,van_der_aalst_federated_2021,van_der_aalst_practitioners_2022,munoz-gama_event_2022,milde_enabling_2023,berti_generic_2023,pajic_simovic_towards_2021,olveczky_object-centric_2019,di_francescomarino_extracting_2019}. This indicates the difficulty in defining and identifying cases within the data. The literature highlights that case IDs are not always readily available, and defining a case notion requires a deep understanding of both the data and the process.

\textbf{Ch6: Granularity} was the most prominent challenge in event abstraction, with 13 references (22\%) \cite{montali_bermuda_2023,de_smedt_llms_2024,olveczky_object-centric_2019,fahland_event_2020,umer_data_2022,southier_systematic_2023,diba_extraction_2020,de_weerdt_foundations_2022,andrews_leveraging_2019,prostean_event_2020,daniel_pre-hospital_2019,seiger_towards_2020,di_francescomarino_analytics_2023}. This reflects the difficulty in determining the appropriate level of detail for event logs. The literature notes that event data is often recorded at different levels of granularity, making it challenging to abstract events into meaningful process activities.

\textbf{Ch7: Effort- and time-consuming}. The most prominent overarching challenge identified was the effort and time-consuming nature of event log generation, referenced in 12 (20\%) studies. \cite{kampik_event_2022,diba_extraction_2020,munoz-gama_rethinking_2022,benvenuti_interactive_2022,van_der_aalst_practitioners_2022,ceravolo_extracting_2020,gonzalezlopez_connecting_2019,de_smedt_llms_2024,martinez_lagunas_analysis_2024,cabanillas_supporting_2023,marrella_towards_2022,andrews_quality-informed_2020} This highlights the significant manual effort required in the process, often due to the lack of structured methodologies. The literature emphasizes that generating event logs is a labor-intensive task, requiring substantial human involvement, particularly in data extraction and transformation.

\textbf{Ch8: Data convergence / divergence} was mentioned in 11 primary studies (19\%) \cite{diba_extraction_2020,indulska_extracting_2023,prostean_event_2020,munoz-gama_event_2022,montali_virtual_2023,van_der_aalst_federated_2021,berti_generic_2023,kopke_towards_2023,pajic_simovic_towards_2021,ceravolo_extracting_2020,olveczky_object-centric_2019}. Data convergence occurs when the same event relates to multiple cases, leading to duplication in the event log, while data divergence occurs when multiple instances of the same activity are recorded within a single case, making it difficult to correlate events correctly.

\subsection{Data-related requirements for event log generation for process mining (RQ2)}
In this section, we report the requirements that were mentioned in the primary studies and which we have grouped into five large groups. We observed that some primary studies mentioned approach-, tool-, or standard-specific requirements. We report such requirements along with general data-related requirements but correspondingly mention an approach, tool, or standard (XES (eXtensible Event Stream), OCEL (Object-Centric Event Logs)) in the context of which requirements were mentioned.

\textbf{R1: Data elements availability} was one of the most frequently identified data-related requirements and refers to the availability and quality of the fundamental data elements required for process mining, as outlined below.

\textit{R1.1: Timestamps}. (26 studies) Timestamps are crucial for establishing the order of events within a case. Each event must have at least one timestamp, and in some cases, two timestamps (start and end) are preferred for more detailed performance analysis. The timestamps must be precise and consistent to avoid issues with event ordering (e.g., ~\cite{de_weerdt_foundations_2022,fahland_event_2020,marrella_towards_2022,bekeneva_approach_2020,hernandez_semi-automated_2022,indulska_extracting_2023,alzhrani_process-aware_2024,bano_database-less_2021,husin_process_2023,butt_behavioral_2023,tariq_time_2022,geeganage_text2el_2022,schuh_data_2020,umer_data_2022,bulander_development_2023,van_der_aalst_practitioners_2022,montali_deriving_2023,benvenuti_interactive_2022,desjardins_enabling_2024,martinez_lagunas_analysis_2024,di_francescomarino_analytics_2023,di_francescomarino_extracting_2019,sim_automatic_2022}).

\textit{R1.2: Case ID} (23 studies) A critical requirement is the definition of a case identifier, which links multiple events to a single process instance. This is essential for organizing events into coherent sequences that represent individual process executions. The case identifier must be unique and consistent across the dataset to ensure accurate process mining analysis (e.g., ~\cite{de_weerdt_foundations_2022,fahland_event_2020,marrella_towards_2022,schuh_data_2020,umer_data_2022,hernandez_semi-automated_2022,indulska_extracting_2023,alzhrani_process-aware_2024,prostean_event_2020,bano_database-less_2021,montali_deriving_2023,benvenuti_interactive_2022,desjardins_enabling_2024,van_der_aalst_practitioners_2022,sim_automatic_2022,just_collaborative_2023,butt_behavioral_2023,tariq_time_2022,geeganage_text2el_2022,di_francescomarino_extracting_2019,bulander_development_2023,husin_process_2023,montali_virtual_2023}). 

\textit{R1.3: Activities} (19 studies) (specific steps in the process) Each event in the log must correspond to an activity, which represents a specific step in the process. Activities should be clearly defined and labeled in a way that is understandable to business experts. The granularity of activities should be appropriate for the analysis, capturing meaningful steps without being overly detailed or abstract (e.g., ~\cite{de_weerdt_foundations_2022,bekeneva_approach_2020,schuh_data_2020,hernandez_semi-automated_2022,benvenuti_interactive_2022,desjardins_enabling_2024,sim_automatic_2022,bulander_development_2023,martinez_lagunas_analysis_2024,reichert_ontology-driven_2016,alzhrani_process-aware_2024,prostean_event_2020,bano_database-less_2021,montali_deriving_2023,di_francescomarino_extracting_2019,geeganage_text2el_2022,butt_behavioral_2023,husin_process_2023,montali_virtual_2023}).

\textit{R1.4: Events} (10 studies) Events are the individual occurrences within a process, and each event must be linked to a case and an activity. Events should be well-defined and mapped to their respective cases to ensure accurate process representation\cite{de_weerdt_foundations_2022,marrella_towards_2022,umer_data_2022,tariq_time_2022,hernandez_semi-automated_2022,indulska_extracting_2023,alzhrani_process-aware_2024,prostean_event_2020,bulander_development_2023,martinez_lagunas_analysis_2024}.

\textit{R1.5: Labels} (10 studies) Labels provide additional context or categorization for events and activities, and assist in distinguishing between different types of events or activities (e.g., \cite{de_weerdt_foundations_2022,hernandez_semi-automated_2022,bekeneva_approach_2020,schuh_data_2020,indulska_extracting_2023,bano_database-less_2021,benvenuti_interactive_2022,desjardins_enabling_2024,sim_automatic_2022}).

\textit{R1.6: Additional attributes} (e.g., resources involved, costs) which provide context or supplementary information and can be required for process analysis, application of filtering, comparative analysis, or deeper insights into the process \cite{de_weerdt_foundations_2022,husin_process_2023,bekeneva_approach_2020,desjardins_enabling_2024,montali_virtual_2023,martinez_lagunas_analysis_2024,pajic_simovic_towards_2021}.

\textit{R1.7: Relationships} between events, activities, and cases were mentioned as important for understanding the flow of the process. For example, relationships between events and activities help map which activities are associated with which events, while relationships between cases and events ensure that events are correctly grouped by process instances \cite{fahland_event_2020,umer_data_2022,hernandez_semi-automated_2022}.

\textbf{R2: Data quality attributes}
Multiple primary studies mentioned specific data quality attributes that should be assured.
\textit{R2.1: Accuracy}: The data should accurately reflect the process, with no errors or inconsistencies in the recorded events \cite{andrews_quality-informed_2020}.
\textit{R2.2: ID uniqueness}: Each event and case identifier should be unique to avoid duplication and ensure accurate analysis \cite{andrews_quality-informed_2020}.
\textit{R2.3: Completeness}: The log should not have missing data, especially for critical attributes like case ID, activity, and timestamp\cite{andrews_quality-informed_2020,prostean_event_2020}.
\textit{R2.4: Trustworthiness}: The data should be reliable and trustworthy, meaning that the recorded events actually occurred and their attributes are correct \cite{prostean_event_2020}.
\textit{R2.5: Relevance}: The data should be relevant to the process being analyzed, with no extraneous information that could distort the results \cite{montali_deriving_2023}.
\textit{R2.6: Consistency}: The data should be consistent in terms of timestamps, data types, and case IDs. Consistency ensures that the log can be accurately analyzed without errors caused by mismatched or inconsistent data \cite{montali_deriving_2023,van_der_aalst_practitioners_2022}.
\textit{R2.7: Reliability}: The data should be reliable, meaning that it can be consistently used for process mining without unexpected errors or issues \cite{arpasat_applying_2021}.
\textit{R2.8: Granularity Level}: The granularity of logging should be appropriate for the analysis, capturing enough detail to be meaningful but not so detailed that it becomes cumbersome. The granularity level should be understood and interpreted by business experts to ensure the usefulness for analysis \cite{de_weerdt_foundations_2022}.

\textbf{R3: Data access and structure}
\textit{R3.1: Assuring access to data.} Access to the source data is essential for extracting and transforming the data into event logs. The data should be accessible and well-structured to facilitate the extraction process \cite{alzhrani_process-aware_2024,bano_database-less_2021}. This also includes the ease of identifying the relevant data elements~\cite{diba_extraction_2020}.

\textit{R3.2: Assuring appropriate data structure}  The database structure must be well-defined and understood to ensure accurate data extraction and transformation \cite{hernandez_semi-automated_2022,alzhrani_process-aware_2024,bano_database-less_2021}. This can include (1) the database schema and/or mapping to capture the relationships between objects, events, and attributes~\cite{diba_extraction_2020,cabanillas_supporting_2023} (Redo-Log Based/Database-Less, object-centric approaches), ensuring accurate representation of the process\cite{diba_extraction_2020} and accurate data extraction and transformation \cite{bano_database-less_2021,hernandez_semi-automated_2022}, (2) the presence of a primary key constraint column for identifying unique records in the database (RDB2Log Approach) \cite{andrews_quality-informed_2020}, (3) foreign key constraints columns to help define relationships between tables for accurate event log generation\cite{andrews_quality-informed_2020} (RDB2Log Approach).

\textbf{R4: Clear definitions and mapping}
\textit{R4.1: Specify clear definitions} 
A number of core definitions and terms should be defined to support process mining activities (e.g., the case definition, how cases are defined, should be carefully selected to ensure that events are correctly grouped \cite{marrella_towards_2022}). In addition, definitions need to be understandable to the stakeholders involved in process mining. For example, activities should be labeled in a way that is meaningful and understandable to business experts \cite{de_smedt_llms_2024}. The other definitions include, but are not limited to events~\cite{marrella_towards_2022}, activity names~\cite{marrella_towards_2022}, event types~\cite{marrella_towards_2022}, required timestamp characteristics~\cite{marrella_towards_2022}.

\textit{R4.2: Events Categorization} Different event types should be categorized and labeled appropriately \cite{marrella_towards_2022}.

\textit{R4.3: Mapping of Events to Activities/Cases} Events must be clearly defined and mapped to their respective cases and activities \cite{marrella_towards_2022,diba_extraction_2020}. Mappings are also essential to define how data elements map to event log components\cite{diba_extraction_2020} and to ensure accurate process representation.

\textit{R4.4: Verification of definitions} Definitions should be verified to be accurate and consistent. Primary studies especially emphasize the importance to verify timestamps\cite{marrella_towards_2022}.

\textbf{R5: Capturing of Domain Knowledge} An understanding of the domain and the underlying business processes is essential for accurately defining events, activities, and case notions. \cite{hernandez_semi-automated_2022} suggested that the semantics of the data (i.e., the meaning of the data elements) must be clearly defined to ensure accurate mapping to event log components. 17\% of the primary studies mention embedding domain knowledge in some forms as a requirement \cite{umer_data_2022,hernandez_semi-automated_2022,seiger_towards_2020,montali_deriving_2023,moctar_mbaba_blockchain_2022,milde_enabling_2023,fahland_event_2020,di_francescomarino_analytics_2023,pajic_simovic_towards_2021,gonzalezlopez_connecting_2019}. Domain knowledge can be captured in a form of (1) definitions/vocabulary mapping~\cite{diba_extraction_2020} (XESame) in which terms used in the data are mapped to terms used in process model and business operations, and/or (2) specification of contextual information~\cite{fahland_event_2020} or rules (OBDA Approach)~\cite{diba_extraction_2020} required to ensure that the data is interpreted correctly within the context of the process, (3) domain model (OBDA and XESame approaches) or process model~\cite{cabanillas_supporting_2023} is required to define the structure and relationships of the data\cite{diba_extraction_2020} or to enable natural language processing~\cite{cabanillas_supporting_2023}, (4) domain ontology (OBDA Approach) required to define the relationships between data elements and process concepts \cite{diba_extraction_2020}. \cite{fahland_event_2020} suggested the application of abstraction tables to aggregate and abstract event data, ensuring that the resulting logs are suitable for higher-level process analysis \cite{fahland_event_2020}.

\scriptsize
\begin{longtable}[ht!]
{|p{3.25cm}|p{1.25cm}|p{7.12cm}|p{4.62cm}|}
\hline
\rowcolor{headerbg}
\textbf{Challenges} & \textbf{\# / \% of studies} & \textbf{Challenge description} & \textbf{Requirements} \\ \hline
\endfirsthead
\hline

\endhead
Ch1: Data quality issues
& 23 (39\%)
&  & R1: Data elements availability, R2: Data quality attributes \\ \hline

Ch1.1: Missing data
& 12 (20\%)
& Event logs may be incomplete, with missing events, attributes, or case IDs. & R2.3: Data completeness \\ \hline

Ch1.2: Inconsistent data
& 9 (15\%)
& Inconsistencies in how data is recorded across systems or over time can lead to errors in event logs. & R2.4: Data trustworthiness, R2.6: Data consistency \\ \hline

Ch1.3: Imprecise data
& 6 (10\%)
& Data attributes (often timestamps) recorded at too coarse or inconsistent granularity, affecting ordering or analysis. & R2.1: Data accuracy; R2.7: Data reliability \\ \hline

Ch1.4: Incorrect data
& 5 (8\%)
& Data may be recorded incorrectly, leading to inaccurate event logs. & R2.4: Data trustworthiness, R2.7: Data reliability \\ \hline

Ch1.5: Irrelevant data
& 3 (5\%)
& Data that is not relevant to the process may be included in event logs, leading to noise & R2.5: Data relevance \\ \hline

Ch2: Data integration
& 16 (27\%)
& Integrating data from multiple sources into a coherent event log is challenging, particularly when formats vary. & R3: Data access and structure, R4: Clear definitions and mappings \\ \hline

Ch3: Non-process centric data 
& 14 (24\%)
& Data that is not recorded with a process perspective requires significant transformation. & R4: Clear definitions and mappings, R5: Capturing of domain knowledge \\ \hline

Ch4: Heterogeneity
& 13 (22\%)
& Data is often stored in different formats and structures across multiple systems, making integration difficult. & R3: Data access and structure, R4: Clear definitions and mappings, R5: Capturing of domain knowledge \\ \hline

Ch5: Case notion
& 13 (22\%)
& Defining and identifying cases within the data is challenging, especially when case IDs are not explicitly recorded. & R4: Clear definitions and mappings, R5: Capturing of domain knowledge \\ \hline

Ch6: Granularity 
& 13 (22\%)
& Determining the right level of detail for event logs is challenging, especially with varying data granularity. & R2.8: Granularity level \\ \hline

Ch7: Effort/time-consuming
& 12 (20\%)
& Event logs generation is largely manual and time-consuming, often due to a lack of structured methods. & - \\ \hline

Ch8: Data convergence / divergence
& 11 (19\%)
& The same event may appear across multiple cases, causing duplication in the event log, and repeated activity instances within a case can hinder accurate event correlation. & R4: Clear definitions and mappings \\ \hline

Ch9: Event data identification
& 7 (12\%)
& Identifying relevant data can be challenging, especially
when data is scattered across systems. & R3: Data access and structure, R4: Clear definitions and mappings, R5: Capturing of domain knowledge \\ \hline

Ch10: Process scoping
& 6 (10\%)
& Defining the boundaries of a process (start to end) and determining which events belong to which process instances. & R3: Data access and structure, R4: Clear definitions and mappings, R5: Capturing of domain knowledge \\ \hline

Ch11: Data complexity
& 6 (10\%)
& The complexity of data structures, such as one-to-many or many-to-many relationships, can make extraction difficult. & R3.2: Data structure  \\ \hline

Ch12: Locating data
& 5 (8\%)
& Finding relevant data sources or tables for event log generation, especially when they are scattered or undocumented. & R4: Clear definitions and mappings, R5: Capturing of domain knowledge \\ \hline

Ch13: Documentation
& 4 (7\%)
& Lack of or inconsistent documentation. & R4: Clear definitions and mappings \\ \hline

Ch14: Data volume
& 4 (7\%)
& Handling large datasets can be challenging, especially when integrating data from multiple sources. & R2: Data quality attributes, R3: Data access and structure \\ \hline

Ch15: Ambiguity
& 4 (7\%)
& Event labels may be ambiguous or inconsistent across systems, hindering abstraction into process activities. & R1: Data elements availability, R4: Clear definitions and mappings \\ \hline

Ch16: Naming
& 4 (7\%)
& Labels or attributes are used inconsistently, requiring standardization. & R1: Data elements availability, R4: Clear definitions and mappings \\ \hline

Ch17: Duplicates
& 3 (5\%)
& Events or activities repeated, leading to confusion & R1: Data elements availability, R4: Clear definitions and mappings \\ \hline

Ch18: Data variability
& 3 (5\%)
& Changes in data recording over time or across systems hinder consistent extraction and interpretation. & R1: Data elements availability, R3: Data access and structure \\ \hline

Ch19: Data deficiency
& 3 (5\%)
& Data may be lost when transforming raw data into event logs, especially when flattening complex data structures. & R2.3: Completeness, R3: Data access and structure \\ \hline

Ch20: Loss of information
& 3 (5\%)
& Key relationships or attributes are lost if the log is forced to a single perspective or simplified case notion. & R1: Data elements availability, R2: Data quality attributes \\ \hline

Ch21: Data availability
& 2 (3\%)
& Some needed data might not be available & R1: Data elements availability, R2: Data quality attributes \\ \hline

Ch22: Logging variability
& 2 (3\%)
& Inconsistencies in how events are logged across different systems or over time. & R2: Data quality attributes, R3: Data access and structure \\ \hline

Ch23: Lack of standardization
& 2 (3\%)
& The absence of standardized logging formats. & R2: Data quality attributes \\ \hline

Ch24: Semantic interoperability
& 1 (2\%)
& Need for shared meaning or domain alignment across data sets. & R4: Clear definitions and mappings, R5: Capturing of domain knowledge \\ \hline

Ch25: Impedance mismatch
& 1 (2\%)
& Mismatches between source data structures, ontologies, and process mining tools can hinder event log construction. & R4: Clear definitions and mappings \\ \hline

Ch26: Data collection
& 1 (2\%)
& Collecting data from multiple, often unknown or unstructured sources is challenging, especially without metadata. & R3: Data access and structure, R5: Capturing of domain knowledge \\ \hline

Ch27: Process complexity
& 1 (2\%)
& Complex processes with many activities or decision points can make it difficult to define a clear scope. & R5: Capturing of domain knowledge \\ \hline

Ch28: Process semantics
& 1 (2\%)
& Understanding domain-specific terms or activity definitions to properly interpret events. & R4: Clear definitions and mappings, R5: Capturing of domain knowledge \\ \hline

Ch29: Formatting errors
& 1 (2\%)
& Incorrectly formatted fields. & R1: Data elements availability, R2: Data quality attributes \\ \hline

\caption{Mapping of the identified challenges to main types of requirements}
\label{tab:table}

\end{longtable}
\normalsize

\subsection{Techniques applicable to data-related challenges in event log generation (RQ3)}
In this section we report the identified data engineering and requirements engineering solutions for data quality in process mining.

\subsubsection{Techniques overview}
Three primary studies we have selected~\cite{diba_extraction_2020,de_weerdt_foundations_2022,fahland_event_2020} suggested a categorization of techniques for event log preparation into three groups: techniques for event data extraction (dealing with identification of data elements that characterize events coming from heterogeneous data sources), correlation (focused on grouping the data elements related to a single process instance), and abstraction techniques (looking at the mapping of data elements to events corresponding to activity executions in a business process and aiming to bridge differences in the granularity at which data is recorded). Next, we use these groups to better report the role of data and requirements engineering.

Beyond this categorization there was a wide range of solutions suggested to address data-related challenges and requirements, such as application of process mining event log standards (e.g., XES~\cite{desjardins_enabling_2024}, eXtensible Object-Centric (XOC)~\cite{diba_extraction_2020}, OCEL~\cite{desjardins_enabling_2024}), metrics (e.g., completeness or uniqueness (the proportion of non-unique values in a column) metrics~\cite{andrews_quality-informed_2020}). Among the selected primary studies we have identified one study utilizing Large language Models (LLMs)~\cite{de_smedt_llms_2024} for analyzing the event data, generating meaningful labels, and recommending relevant connectors, based on natural language prompts and queries.

\textbf{Extraction}
Existing extraction techniques aim to identify data elements that characterize events coming from heterogeneous data sources by using ontologies (Ontology-Based Data Access, OBDA), redo logs, database objects~\cite{marrella_towards_2022,andrews_quality-informed_2020}, and/or artifacts or objects~\cite{diba_extraction_2020}. In this case, ontologies, redo logs, data base objects, or artifacts/objects are used for guiding the extraction of the relevant data. For example, Redo-logs-based extraction approach starts by defining an event model which relates database changes to events. In Ontology-Based Data Access (OBDA), data is accessed through an ontology that is linked to the database schema. Artifact-based approaches preserve the object-centric nature of data by considering artifacts as the case identifier, instead of transforming data into a process-oriented view.

Data extraction tools include RDB2Log allowing to flexibly generate high-quality event logs from relational databases on the basis of a tailored data quality assessment~\cite{andrews_quality-informed_2020}, ProM Import Framework~\cite{de_weerdt_foundations_2022}, XESame~\cite{de_weerdt_foundations_2022}, XOC Log Generator is a plug-in in ProM which automatically extracts XOC logs from redo logs or change tables of databases, after relevant event types are specified using domain knowledge~\cite{diba_extraction_2020}, EVS Model Builder supporting the extraction of SAP transaction data, and
Onprom (OBDA-based toolchain and data extraction methodology).

There are also studies trying to automate the process of data extraction, for example, with automated mapping of database schema and process model using natural language processing~\cite{cabanillas_supporting_2023}.

\textbf{Correlation}
Event correlation techniques are applied to group the data elements that relate to a single process instance. As reported in ~\cite{diba_extraction_2020} 
probabilistic approach based on Markov chains and an expectation–maximization technique to correlate events with process instances as well as discovering a process model, 
sequence partitioning, MapReduce-based approach, or manual identification of relevant tables and relations of the database for the selected case notion. This group of techniques is not closely related to our research question.

\textbf{Abstraction}
Event abstraction techniques are aiming to bring lower level events to a better granularity level and map of data elements to events that correspond to activity executions in a business process. Multiple techniques are applicable for abstraction such as techniques for supervised learning, neural networks-based techniques, clustering based on attribute values, temporal proximity and attribute values, pattern matching, learning of compound features that induce abstraction, grouping using linguistic relations and textual, semantic distances, supervised sequence identification of frequent sequential patterns for abstraction, matching of process models and event data based on ordering constraints, combined model-based abstraction with temporal analysis of data, etc.

Next, we provide the results of our analysis about the application and applicability of data engineering and requirements engineering techniques in process mining.

\subsubsection{Data engineering techniques in process mining}
Existing studies report a number of data engineering techniques mainly applied for exporting, and integrating data. Specifically, such techniques and tools widely used in data engineering were mentioned Extract-Transform-Load (ETL) processing and corresponding scripts~\cite{de_weerdt_foundations_2022,indulska_extracting_2023}, Microsoft Power Query~\cite{bulander_development_2023}, \cite{de_weerdt_foundations_2022}, Microsoft SQL server~\cite{munoz-gama_rethinking_2022}, Python for data transformation~\cite{munoz-gama_rethinking_2022}. Rarely, some of the more advanced data engineering techniques were mentioned such as Complex Event Processing~\cite{seiger_towards_2020}, distributed data processing for distributed event log integration~\cite{bekeneva_approach_2020}, or data federation that prevents the creation of yet another duplicated database or data store, but instead provides flexible querying and analysis tools for information from multiple source systems as if all data resides within a single integrated database~\cite{de_weerdt_foundations_2022,van_der_aalst_federated_2021}. It is noteworthy that some studies suggest the demand for approaches or tools to support process analysts in extracting event logs~\cite{andrews_quality-informed_2020}, which indicates a demand to develop tools with data engineering capabilities useful for non-data experts.

\subsubsection{Requirements engineering}
Primary studies we have reviewed have not mentioned the application of requirements engineering techniques for the purposes of process mining directly. However, one of the important challenges related to requirements engineering was application of domain knowledge in event log generation.
Specifically, domain knowledge challenges emerged across all three groups of techniques for event log preparation.

\cite{indulska_extracting_2023} suggest the following types of understanding or domain knowledge required, which are as follows, (1) data understanding (definition of the process mining questions/cornerstones and decision of which relevant tables to use for the procedure), (2) conceptual understanding (identification and documentation of which document/object/relationship types are represented in these tables), (3) perspective, granularity, and scoping (identification of a suitable notion of case and of the relevant events pertaining the elicited activities).
Suggested approach involves both domain experts and data engineers~\cite{indulska_extracting_2023}.

In data extraction some form of high level conceptual view of the domain or involvement of domain experts is required to define the relevant domain concepts and relationships~\cite{indulska_extracting_2023}, understand the available data~\cite{umer_data_2022}, determine relevant data sources~\cite{fahland_event_2020}, identify the relevant data in those sources~\cite{marrella_towards_2022,berti_generic_2023}, obtain the relations of the data elements across different sources, and map event log elements to available data~\cite{pajic_simovic_towards_2021}. For example, XOC Log Generator requires specification of relevant events using domain knowledge. The role of domain knowledge for specification of the concepts related to extracted data is even more important in the case of cross-system process mining~\cite{just_collaborative_2023}.

Majority of the abstraction techniques claim no demand for domain knowledge or involvement of domain experts, and applicability of unsupervised learning techniques, they still often leverage additional domain knowledge, a process model, annotated traces, or other information~\cite{diba_extraction_2020}.

Event correlation techniques often assume the availability of extra information in addition to raw events, to guide the correlation~\cite{diba_extraction_2020} (e.g., in ~\cite{seiger_towards_2020}), and hence do not analyze the demand for domain knowledge in depth. However, domain knowledge is also required for many of these approaches. \cite{fahland_event_2020} suggests that it is ``less a technical question of how to correlate events but, instead, what the most suitable notion of a case is'', which also points to the role of domain knowledge in correlation process.

Some of the log generation techniques are specifically developed to facilitate the involvement of domain experts, such as, domain-specific modeling languages~\cite{valencia-parra_enabling_2019} for extracting event data from transactional databases to enable domain experts to model data of interest at a conceptual level and then automatize the extraction of event logs from ERP systems~\cite{pajic_simovic_towards_2021}, annotation of domain by domain expert using DL-Lite~\cite{reichert_ontology-driven_2016}, to capture and maintain the link between domain knowledge recorded by domain expert and the data in the information system (in BERMUDA method)~\cite{montali_bermuda_2023} or development of domain ontologies~\cite{indulska_extracting_2023}. One of the benefits of methods involving domain experts is maintaining event logs over time when changes happen in the domain or system, because the information is documented consistently in one place~\cite{montali_bermuda_2023}.

Some methods also envisage the evaluation of mapping or automatically generated results by a domain expert~\cite{bano_database-less_2021} or evaluate their automated methods towards domain knowledge~\cite{di_francescomarino_analytics_2023}.

\section{Discussion}\label{sec:discussion}
Our study identified a wide range of 29 challenges and 5 main groups of requirements that address practically all of the challenges. The only exception was Ch7: effort- and time-consuming nature of event log generation that was not addressed with specific requirements. Herewith, we suggest that practically all challenges in event log generation are well-known, and requirements for addressing them are established.

Most often, classical data-related challenges were mentioned, such as basic quality of data or missing data elements (Ch1.1-Ch1.5), characteristics of the data (Ch14-Ch20). The challenges are addressed with quite developed requirements of R1: Data elements availability, and R2: Data quality attributes. Such requirements are well-discussed in the literature; however, there is a lack of guidance on how such requirements can be effectively contextualized in different domains or in specific projects.

Many of the challenges are specific to data processing and engineering activities (e.g., Ch2: Data integration, Ch12: Locating data), which are addressed with R3: Data access and structure requirements; however, in some cases, R4: clear definitions and mappings are demanded to assist in data engineering activities.

Another distinctive group of challenges is related to the domain knowledge required in the process of process mining (e.g., Ch3: Non-process-centric data, Ch5: Case notion). According to our mapping, these challenges can be addressed mainly with R4: Clear definitions and mappings, but in some cases can also be captured with in-depth domain knowledge or involvement of domain experts (R5).

While the identified challenges are specific to process mining, some of them are mainly stemming from the way logging is implemented in software systems, for example, Ch1 (Ch1.1: Missing data), Ch3: Non-process centric data, Ch6: Granularity, Ch22: Logging variability, Ch23: Lack of standardization, Ch29: Formatting errors. Such challenges can be mainly addressed through the implementation of data quality-related requirements (R1, R2). As we suggested, providing some recommendations to software developers on data elements (R1) and guidance on the relevant quality attributes (R2) can be an approach to resolve these challenges.

It is noteworthy that some of the challenges are mapped to two or three requirements, which shows the close interconnection in fulfillment of such requirements for even log generation purposes. For example, process scoping (Ch10) requires both data access and structure (R3), clear definitions and mappings (R4), and some domain knowledge (R5).

This close interconnection between data engineering and domain knowledge processing is also confirmed by the results of our analysis of existing solutions. Many of these solutions and techniques (re)use data engineering techniques for data extraction and try to address domain knowledge challenges to which requirements engineering techniques could be applicable. Next, we discuss in more detail the potential contribution. 

We suggest that data engineering methods can play a fundamental role in addressing some of the challenges to even log generation of process mining. Specifically, advanced data engineering techniques like data federation can be instrumental in resolving data integration challenges in event log generation.

The potential contribution of requirements engineering is twofold. First, system behavior logging implementation mostly depends on the knowledge and expertise of software developers~\cite{rong2017systematic}, and requirements engineering can assist in a synthesis of best practices for the implementation of logging functionality in software systems and assure quality of event logs at their core. Among the primary studies we selected, \cite{montali_deriving_2023} suggested a related approach to address even log generation in legacy systems by analyzing source code and identifying code fragments relevant for event log generation. Otherwise, the idea of addressing the data-related challenges and requirements in the process of systems development mainly stays overlooked in process mining research. Secondly, we suggest that requirements engineering methods can be used to capture the initial process mining requirements and scope the required log generation activities. Some of the requirements engineering techniques can be used to capture and model the domain knowledge required throughout the event log preparation activities. Here, utilization of requirements engineering-based approaches is more beneficial in comparison to other standalone modeling techniques, as requirements engineering is not only focused on formulating the initial requirements and capturing the domain knowledge, but also provides the techniques for requirements verification and validation, and provides the basis for quality assurance.

So far, our study has identified only one primary study focusing on the application of LLMs. In the future, the application of this technology can specifically contribute to resolving some of the challenges to event log generation for process mining, including challenges related to domain knowledge availability.

\section{Threats to validity}\label{sec:threatsValidity}
In order to address threats to validity, we followed the available guidelines for systematic literature reviews. Both the first and the second authors of the study have reviewed and discussed the results of keyword and search query trials. We have applied a set of gold standard papers that we looked for in the search results. In the process of data extraction, the first and second authors of the study repeatedly discussed the data extraction process. We also executed a trial snowballing to identify the need for snowballing. During the data analysis, both authors repeatedly discussed the intermediary results. Finally, analysis of the data was conducted in two rounds. In the first round, the first analysis and categorization were conducted by the first author, and during the second round of the analysis, the second author conducted his analysis and suggested changes to the initial analysis results. The mapping between challenges and requirements (see Table\ref{tab:table}) was carried out based on the authors’ interpretation, as the primary studies lacked consistent and explicit descriptions of both aspects in their interconnection. To mitigate potential threats to validity arising from this subjectivity, the first and second authors independently reviewed the challenges and requirements across the selected studies and performed the mapping in two iterative rounds. During this process, discrepancies and ambiguities were carefully discussed and resolved. To enhance transparency and reproducibility, we provide an open dataset that documents the mapping between challenges and requirements as reported in the studies. Furthermore, we encourage future researchers to adopt more consistent and explicit reporting practices when discussing challenges and requirements in event log generation.

\section{Conclusion}\label{sec:conclusion}
Event log generation is one of the initial and fundamental phases in process mining. Its importance is determined by its role in processing raw event data and system event logs into business event logs applicable for process mining. Due to the importance of this phase of process mining, we completed our systematic review of data-related challenges and requirements for event log generation. The wide variety of challenges reported in the literature can be addressed through the implementation of five basic groups of requirements.

Also, we have analyzed the available solutions for event log generation and identified that data engineering and domain knowledge engineering techniques are already applied in some of the available solutions; however, not in a systematic way. Our analysis reveals that data engineering techniques have the potential to contribute to the resolution of extraction and integration challenges. Requirements engineering techniques have the potential to contribute to addressing the domain and knowledge engineering challenges important for scoping, mapping, and analyzing the event logs. Requirements engineering can also contribute to log event generation by guiding software engineers in developing systems implementing logging functionality in a way best suited for event log generation and process mining.

We suggest that process mining research and practice would closer explore the intersections with the adjacent areas of research and practice, such as requirements engineering, responsible for the engineering of the domain and knowledge of system operation, software engineering, responsible for the implementation of logging functionality and quality of the produced logs, and data engineering responsible for processing and integrating the data required for process mining.

\bibliographystyle{splncs04}
\bibliography{bibliography}

\end{document}